\documentclass[twocolumn,preprintnumbers,superscriptaddress]{revtex4-2}

\usepackage{hyperref}
\usepackage[utf8]{inputenc}

\usepackage{soul}
\usepackage{url}

\usepackage{braket}
\usepackage{amsmath}
\usepackage{dsfont}
\usepackage[english]{babel}
\usepackage{graphicx} 
\usepackage{tikz}
\usetikzlibrary{quantikz}

\usepackage{hyperref}
\usepackage{natbib}
\usepackage{physics}

\graphicspath{ {./images/} }

\newcommand{\DSD}{
Engineering Department
Research Group on Data Science for the Digital Society
La Salle - Universitat Ramon Llull
Carrer de Sant Joan de La Salle, 42
08022 Barcelona (Spain)
}
\newcommand{\UVA}{
Universidad de Valladolid
C/Plaza de Santa Cruz, 8, 
47002 Valladolid (Spain)
}
\newcommand{\LDIG}{
Lighthouse Disruptive Innovation Group, LLC
7 Broadway Terrace, Apt 1
Cambridge MA 02139
Middlesex County, Massachusetts (USA)
}
\newcommand{\orcid}[1]{\href{https://orcid.org/#1}{\includegraphics[width=8pt]{orcid.png}}}

\begin{document}
\title{Quantum Enhanced Filter: QFilter}
\author{Parfait Atchade-Adelomou}
\affiliation{\DSD}
\email{parfait.atchade@salle.url.edu}
\affiliation{\LDIG}
\author{Guillermo Alonso-Linaje}
\affiliation{\UVA}
\email{guillermo.alonso.alonso-linaje@uva.alumnos.es}

\begin{abstract}
Convolutional Neural Networks (CNN) are used mainly to treat problems with many images characteristic of  Deep  Learning. In this work, we propose a hybrid image classification model to take advantage of quantum and classical computing. The method will use the potential that convolutional networks have shown in artificial intelligence by replacing classical filters with variational quantum filters. Similarly, this work will compare with other classification methods and the system's execution on different servers.
The algorithm's quantum feasibility is modelled and tested on Amazon Braket Notebook instances and experimented on the Pennylane's philosophy and framework.
\end{abstract}
\date{March 2021}

\maketitle

\section{Introduction}
From a classical point of view, artificial intelligence has appeared as a solution to some problems that until then had been very difficult to deal with, among which we can highlight the classification of images. The appearance of neural networks, and specifically Convolutional Neural Networks (CNN)\cite{Albawi2017,oshea2015introduction,kalchbrenner2014convolutional}, introduced a significant improvement in this task. Throughout this article, we will try to show a possible application of current quantum computers in such classification tasks by creating a hybrid model and classical convolutional networks that have already demonstrated their potential in this field.

The emerging field of hybrid quantum-classical algorithms joins CPUs and QPUs\cite{Karalekas2020} to speed up specific calculations within a classical algorithm. This allows for shorter quantum executions that are less susceptible to the cumulative effects of noise and run well on today's devices. This is why we intend to explore the performance of a hybrid convolutional neural network model that incorporates a trainable quantum layer by one hand, and by the other, effectively replacing a convolutional filter in both quantum simulators and QPU.

We propose to design a trainable quantum convolutional filter in a hybrid neural network, appealing for the NISQ era, inspired by these papers \cite{liu2019hybrid, henderson2019quanvolutional}, but generalizing these previous works, and using cloud-based QPU.

\section{Related Work}
Since Alan Turing demonstrated in 1936 that there exist non-computable problems, the interest in creating new ways of solving them has grown remarkably. This, together with the consequences of the well-known Moore's Law, gave way to the idea of building quantum computers. Throughout these last decades, the superiority of these new computers has been demonstrated to solve some specific problems such as the factorization of prime numbers through Shor's algorithm\cite{shor1994algorithms} or the search in disordered sets with Grover's algorithm\cite{grover1996fast}, although all this limited to the number of qubits available. We are currently in the NISQ\cite{Joh18} era in which we have computers between 50-100 qubits, opening the way to the emerging field of hybrid quantum-classical computing. Within this, different algorithms have been developed, such as VQE\cite{Dao19}, QAOA\cite{farhi2014quantum} or, in which we will focus, Quantum Machine Learning (QML)\cite{Mar14,Mar19,JBi17,Adr20,adelomou2020using,atchadeadelomou2021quantum}.

Since the emergence of deep learning, CNN has helped accelerate image processing, NLP, or even chemistry; the use of CNN has flooded almost every industry.
The Ref.\cite{schutt2017schnet} proposes using continuous filter convolutional layers to model local correlations without requiring that the data be in a grid. They applied a new deep learning architecture that models quantum interactions in molecules. The following Ref.\cite{Oh2020} verifies whether the QCNN model can efficiently learn compared to CNN through training using the MNIST dataset through the TensorFlow Quantum platform. While Ref.\cite{Cong_2019} provides a generic framework for simultaneously encoding and decoding procedures. The said framework helps to find that the significant schema surpasses known quantum codes of comparable complexity.

In Ref.\cite{henderson2019quanvolutional} the author presents a new type of transformational layer called quantum convolution that operates on the input data by locally transforming the data using a series of random quantum circuits, in a similar way to the transformations performed by layers of Random convolutional filters. Although his work showed that QNN models had higher test set accuracy and faster training compared to purely classical CNNs, it is worth saying that this approach uses the fixed quantum filter, not variational. And what it does is pre-process the data before applying it to the network.

After analyzing the state of the art in CNN deeply, we did not find any approach like the one we propose. Create a variational quantum filter (QFilter), taking advantage of all the classical CNN’s experience and only substitute the scalar product for a quantum one. What we intend with this proof of concept is to make sense of the hybrid computing platform and align with the Pennylane philosophy- Training a quantum computer the same way as a neural network.

\section{CNN}
Convolutional neural networks \cite{oshea2015introduction,kalchbrenner2014convolutional} are used mainly to treat problems with a large number of images characteristic of Deep Learning. Regarding its operation, it could be divided into two stages. The first one will be in charge of passing the image through some filters to create new images that facilitate understanding the network. After this, we will give the previous phase's output through a full-connected neural network capable of learning thanks to a cost function. Finally, it is worth mentioning that both the layers of the neural network and the filters are parameterized, so it is expected that, with the training process, they will be updated towards the desired values.

\subsection{Convolutional filters}
\begin{figure}[htbp]
\centering
\includegraphics[width=0.4\textwidth]{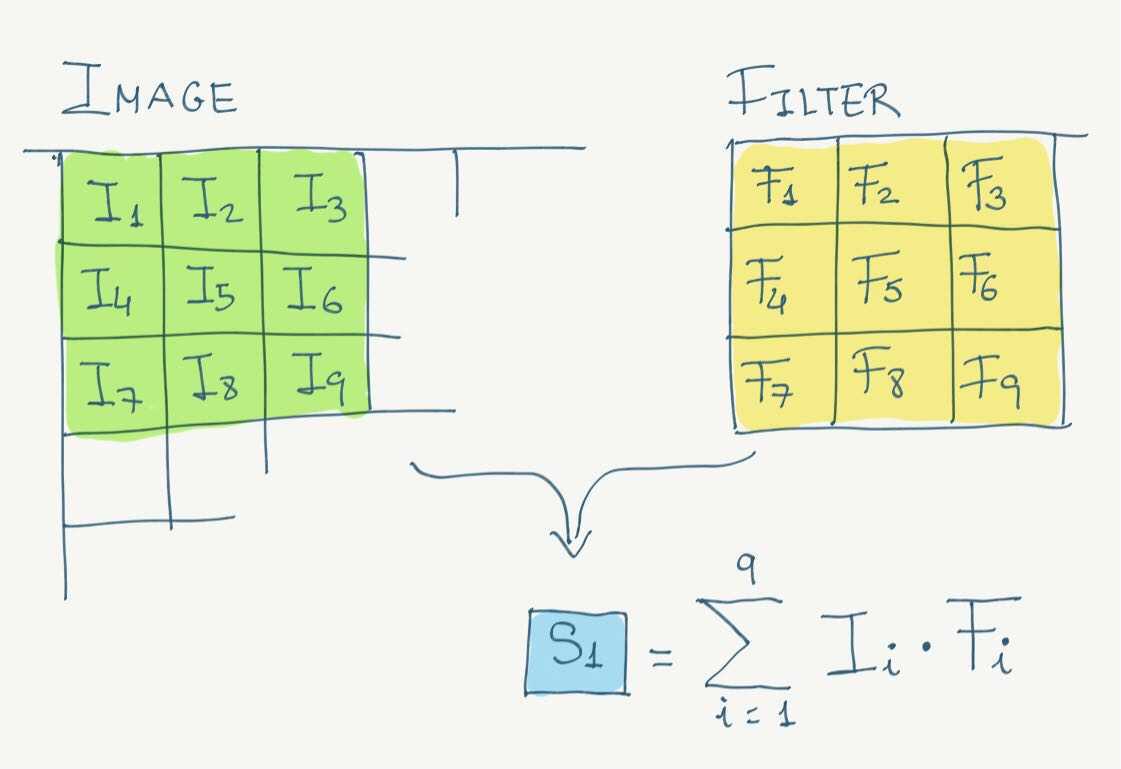}
\centering
\caption{Suppose we take $ n = 3 $, that is, the filter will have a total of $ 9 $ elements. The result after performing the operation is to obtain $ \sum_ {i = 1}^{9} I_ {i} F_ {i} $}
\centering
\label{fig:QFilter}
\end{figure}

Focusing on the first of the phases described, we will talk about convolutional filters. The process followed to transform the image with these follows a simple process. Initially, a filter is defined as an $ n \times n $ matrix, then a window of the same dimension will go through the image performing the operation shown in Fig.\eqref{fig:QFilter}.

As the filter is moved, the solution image's size will be reduced compared to the initial input. There are padding techniques with which it could be possible to make the final image retain its size, but we have chosen not to carry out this process to reduce the number of parameters. In this way, given an image of dimension $l \times w$ and a filter of $n \times n$, we will obtain an output of $[\frac{l}{n}] \times [\frac{w}{n}]$.

\section{Our model}
In this article, we will create a convolutional network to classify the MNIST dataset\cite{Cohen2017} (set of images with handwritten digits $ 0 $ to $ 9$) and the fashion MNIST dataset\cite{xiao2017fashionmnist}. We will work with a convolutional layer in which we will apply $ 4 $ filters, and later we will connect it to a neural layer whose output will have dimension $ 10 $, one for each digit. The outcome represents each class's probability, and we will say that an image belongs to the class whose probability is more outstanding. In these cases in which we want to approximate a probability, the crossed entropy is applied to calculate the defined error $E$ as follows:

\begin{equation}
\label{Error_eq}
    E = \frac{-1}{M}\sum_{i = 1}^{M} y_i log(\hat{y}_i)
\end{equation}

Where $M$ is the number of classes, $ \hat {y} _i $ is the probability obtained from the class $ i $,  and $ y_i = 1 $ if the label is $ i $ or $ 0 $ otherwise.
Up to this point, the process followed could be interpreted as a classical development applying convolutional networks. However, we have decided to carry out a quantum approximation, replacing, in this case, the classical filters with quantum procedures. The birth of this idea arises when representing the elements of the image and the filter as vectors:

\begin{equation}
\label{I_F_eq}
I := \begin{pmatrix}
I_{1} \\
I_{2} \\
 ...       \\
I_{n}
\end{pmatrix}
\hspace{1cm}
F := \begin{pmatrix}
F_{1} \\
F_{2} \\
 ...       \\
F_{n}
\end{pmatrix}
\end{equation}

In this way, it is easy to realize that the operation carried out is nothing more than the scalar product of said vectors. Therefore, we could denote it according to Dirac's notation as $ \braket {I}{F} $. Looking at it this way, that starts to get a sense of the idea behind filter quantization.

\subsection{Quantum filters}

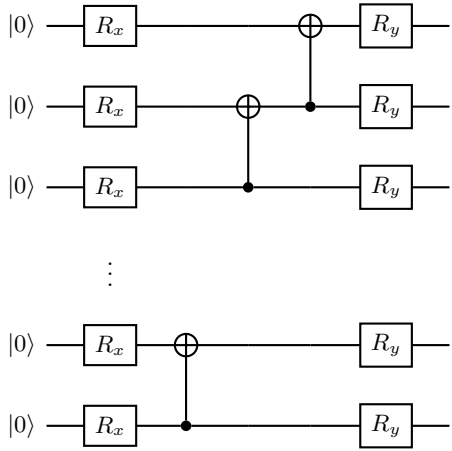
\begin{figure}[htbp]
\begin{center}
\begin{quantikz}
\lstick{$\ket{0}$} & \gate{R_x}{} & \qw & \qw & \targ{} & \gate{R_y}{} & \qw\\
\lstick{$\ket{0}$} & \gate{R_x}{} & \qw & \targ{} & \ctrl{-1} & \gate{R_y}{} & \qw\\
\lstick{$\ket{0}$} & \gate{R_x}{}  & \qw & \ctrl{-1} & \qw & \gate{R_y}{}& \qw\\
&\vdots \\
\lstick{$\ket{0}$} & \gate{R_x}{}  & \targ{} & \qw & \qw &  \gate{R_y}{}& \qw \\
\lstick{$\ket{0}$} & \gate{R_x}{}  & \ctrl{-1} & \qw & \qw & \gate{R_y}{} & \qw
\end{quantikz}
\end{center}
\centering
\caption{Schemes of the ansatz that we use to achieve the quantum convolutional adaptive filter. We use two one-qubit gates $RX$, $RY$ and a $CNOT$ gate to accomplish the entanglement}
\centering
\label{fig:Hadamard_Test2}
\end{figure}

Let us define a quantum filter as that filter that executes the dot product $ \braket {I}{F} $ in a quantum circuit. To transfer this process to the circuit, we must first encode input $I$. By definition, all its elements are numbers between $0$ and $1$ (grayscale), so the most logical embedding is to encode said values at angles with gate $ R_Y $. In this way, the filter size will determine the number of qubits in the circuit, requiring $ n ^ 2 $ qubits. We can then represent the input as $ R_Y (I) \ket {0} ^ n $.
Regarding the filter, we design the ansatz determined by $ n ^ 2 $ parameters shown in Fig.\eqref{fig:Hadamard_Test2}.

For simplicity, we will denote this ansatz as $ F (\theta) \ket {0} ^ n $. Having all this notation, what we want is to obtain the dot product, that is $ \braket {0 ^ n}{I ^ {\dagger} F (\theta) \vert 0 ^ n}$. 

To calculate this product, we are going to make two different approximations. In the first way, we actually compute $\vert \braket {0 ^ n} {I ^ {\dagger} F (\theta) \vert 0 ^ n}\vert ^2$ and consists of constructing the circuit $ \ket {I ^ {\dagger} F (\theta) \vert 0 ^ n} $ and obtaining the probability of measuring $ \ket{0}^n $. In this first scenario, it was decided to calculate the global probability that $\ket{0}^n$ would appear over the rest of the possibilities. Still, it is equivalent (although more efficient in several shots) to calculate the individual probabilities that each qubit takes the value $\ket {0}$ and then multiply all the results obtained.

\begin{figure}
\begin{center}
\begin{quantikz}

\lstick{$\ket{0}$}    & \gate{H}{} & \ctrl{1}                 & \ctrl{1}                  & \gate{H}{} & \meter{}\\
\lstick{$\ket{0}^n$} & \qw\qwbundle{n}        & \gate{I^{\dagger}}{} & \gate{F(\theta)}{}    & \qw

\end{quantikz} 
\end{center}
\centering
\caption{We create a random variable with this circuit whose expected value is the expected real part of a quantum state's observed value concerning $\Re(\braket {0 ^ n}{I ^ {\dagger} F (\theta) | 0 ^ n})$}
\centering
\label{fig:Hadamard_Test}
\end{figure}
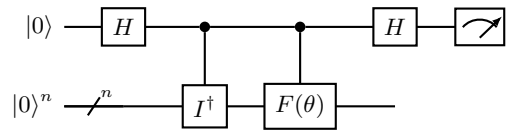

However, although this form may be valid, we will never obtain negative values when calculating the expression's squared modulus when, in fact, the scalar product could have been. For this, we will carry out a second approximation to calculate $ \braket {0 ^ n} {I ^ {\ \dagger} F (\theta) | 0 ^ n} $. In this new scenario, we will calculate the exact value of the inner product (a complex number). However, even if we have the exact value, this does not matter since the neural network will work with real numbers. In this case, what we will do is calculate the real part of said value. For this, we will use the Hadamard Test (see Fig.\eqref{fig:Hadamard_Test}) based on building the following circuit.

In this case, we can define the desired product as shown by equation\eqref{Dot_product}.

\begin{equation}
\label{Dot_product}
    \Re(\braket {0 ^ n}{I ^ {\dagger} F (\theta) | 0 ^ n}) = 2 P(0) - 1
\end{equation}

Where $P (0) $ is the probability of obtaining $ 0 $ when measuring in the previous circuit.

\subsection{Implementation}

To carry out the project, we have taken advantage of the fact that Pennylane \cite{bergholm2020pennylane} provides an interface for Tensorflow and, in this way, use functions already created efficiently. Thanks to this, we have built a hybrid training flow in which we use two different optimizers: SGD \cite{SGD,keskar2017improving} to update the classical parameters and Adadelta \cite{zeiler2012adadelta} for the quantum ones. We have made this distinction since, during the experimentation, we observed that the gradient of the quantum parameters was imposed on the classics without letting that part learn; however, we guarantee double training during the experimentation.

\section{Results}
Before comparing our QFilter in a general way with the classical performance and comparing it with other contributions seen, it is of the utmost importance that we validate its operation globally and affirm that QFilter does meet our expectations and works as we expected. We wanted to analyze the entire test set based on our training accurately through the test benches we made.

\begin{figure}[htbp]
    \centering
    \includegraphics [width=0.4\textwidth]{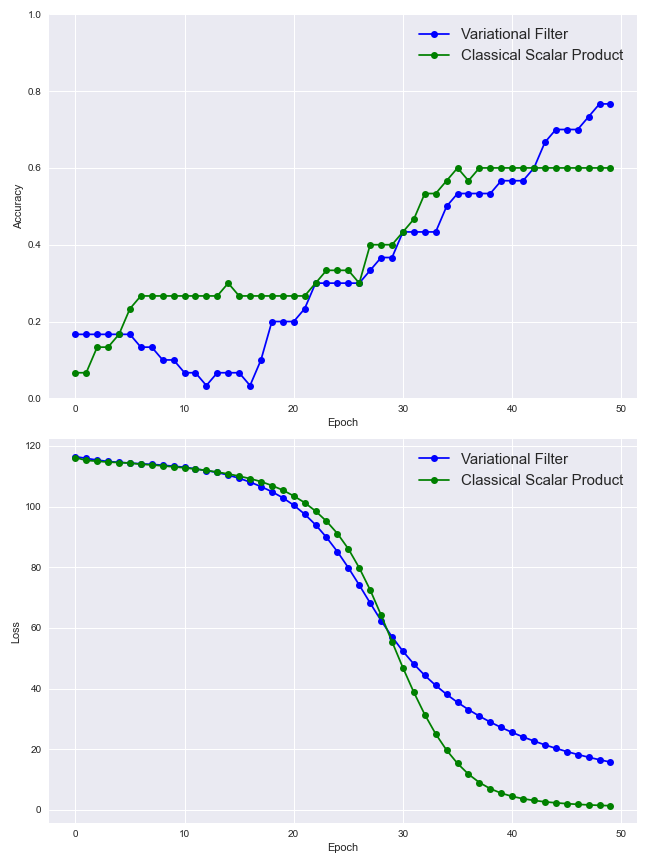}
    \caption{The QFilter results as quantum variational with 50 epochs.}
    \label{fig:Result_QFilter_50_Variat}
\end{figure}

Once the model is built, the first step is to see that our model is capable of learning and generalizing. To do this, we take $50$ images from our dataset (MNIST) and trained it for $30$ and $50$ epochs obtaining the following results Fig.\eqref{fig:Result_QFilter_50_Variat}.

In this case, starting at the $ 15th $ epoch, the model begins to have a precision of $ 40 \% $. It may seem like a bad result on the surface, but let us analyze it. As we said before, we have used $ 50 $ images, and we are trying to divide between a set of $ 10 $ different classes; that is; we are using no more than five images for each class. Despite this, we have run this circuit with a traditional model reaching similar limits as a check.

On the other hand, when implementing the model for the first time, with the filters' size $ 2 \times 2 $, we faced difficulty with the number of operations (dot products). To understand this, let us calculate said number of operations ($ S $) in a generic way for $ p $ image of $ l \times l $ and $ m $ filters of $ n \times n $ during $ t $ epochs.
\begin{equation}
\label{op_eq}
    S =  [\frac {l} {n}] ^{2} m p t
\end{equation}
Therefore, in our initial case of $ 30 $ times $ 4 $ filters of size $ 2 $ and with $ 50 $ images of $ 28 \times $ 28, we performed a total of $ 1,176,000 $ operations. We decided to increase the filter size to $ 4 $ to solve this problem, reducing $ 294,000 $. The reduction in the number of operations is significant, and really what we are doing is transferring the computational load of the classical part to the quantum scalar product because by increasing the size of the filter, we raise the size of the circuit from $ 4 $ qubits to $ 16 $. We were looking for this since we can easily enhance this gradient calculation with the parallelism offered by Amazon Braket with Pennylane API\cite{PennyL-AWS_Braket}.

\begin{figure}[htbp]
\label{Amazon_Braket_fig}
\includegraphics[width=0.4\textwidth]{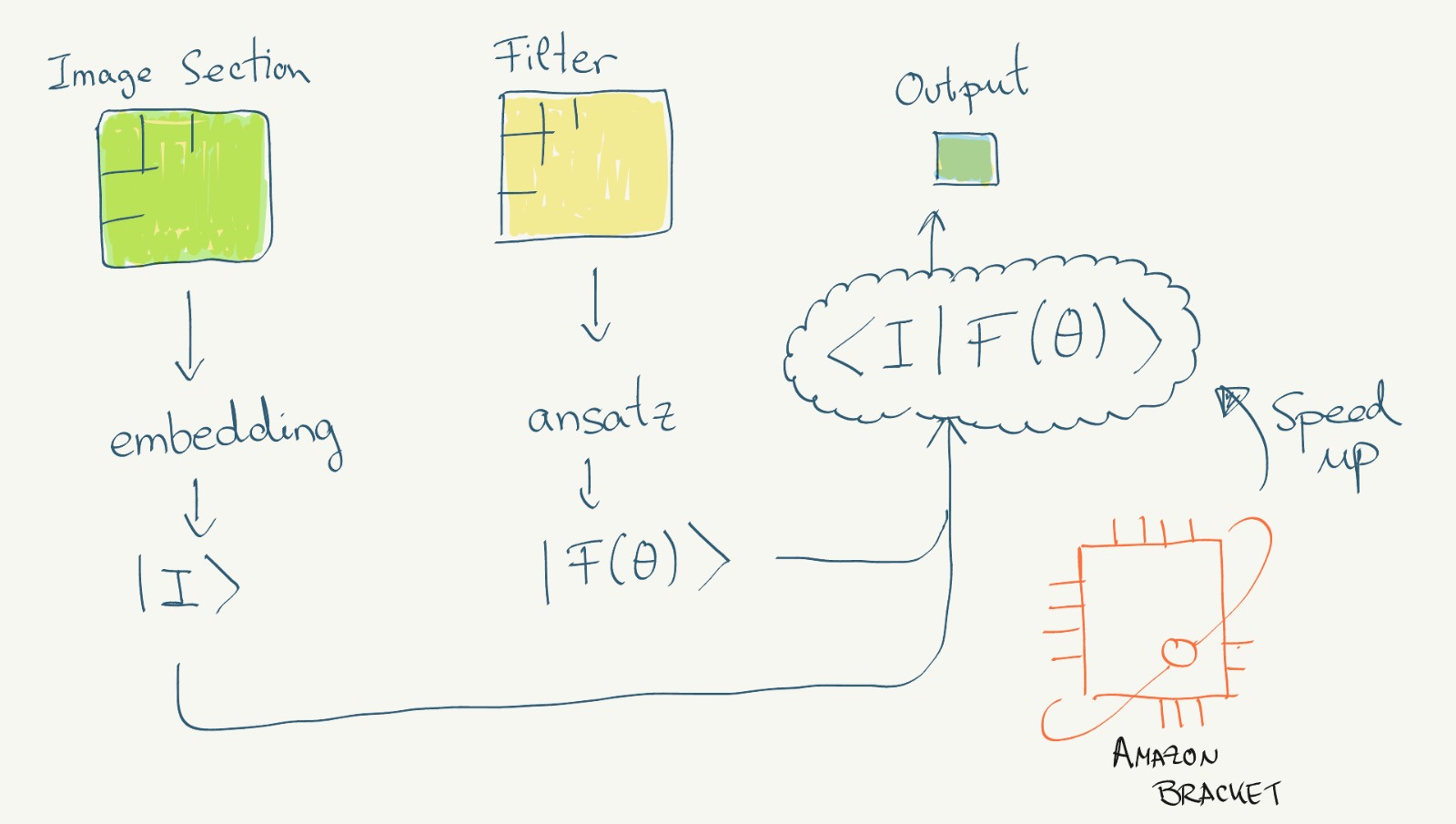}
\centering
\caption{With Amazon Braket we can enhance the quantum scalar product that we had defined as $ \braket {0 ^ n} {I ^ {\ \dagger} F (\theta) | 0 ^ n} $ in parallel. With the API, Pennylane-Braket we can use the high-performance state vector simulator (SV1)\cite{SV1_AWS_Braket} that is designed with parallel execution   to run in parallel all the circuits needed to compute a gradient.}
\centering
\end{figure}

After experimenting with the different test benches, we can present the results. Figure \eqref{fig:Result_QFilter_50_Variat} shows the good behaviour of the variational filter in front of the classical one.

\begin{figure}[htbp]
    \centering
    \includegraphics [width=0.4\textwidth]{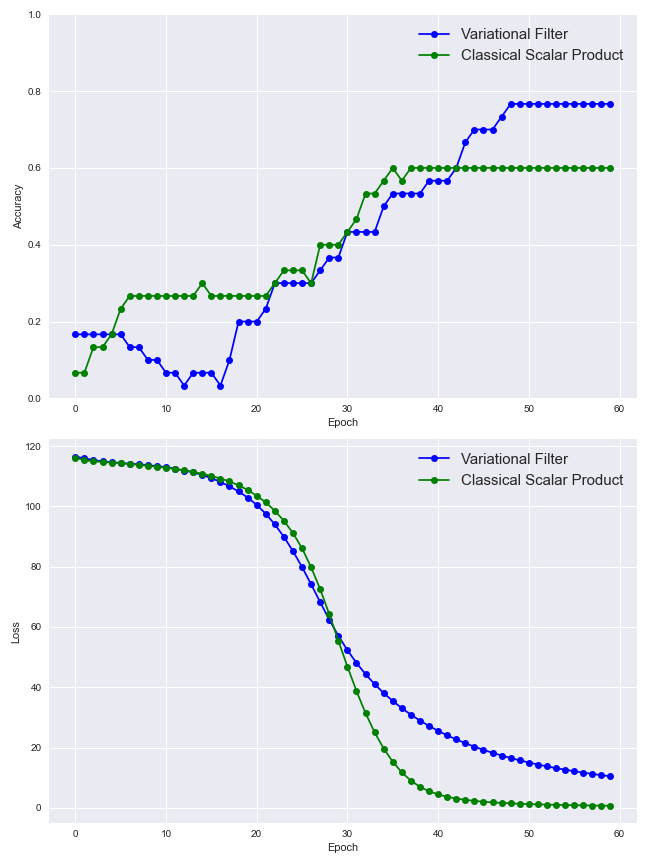}
    
    \caption{The QFilter results as quantum variational with 60 epochs. In this case, SGD is the QFilter's optimizer.}
    \label{fig:Result_QFilter_60_Variat}
\end{figure}

\begin{figure}[htbp]
    \centering
    \includegraphics [width=0.4\textwidth]{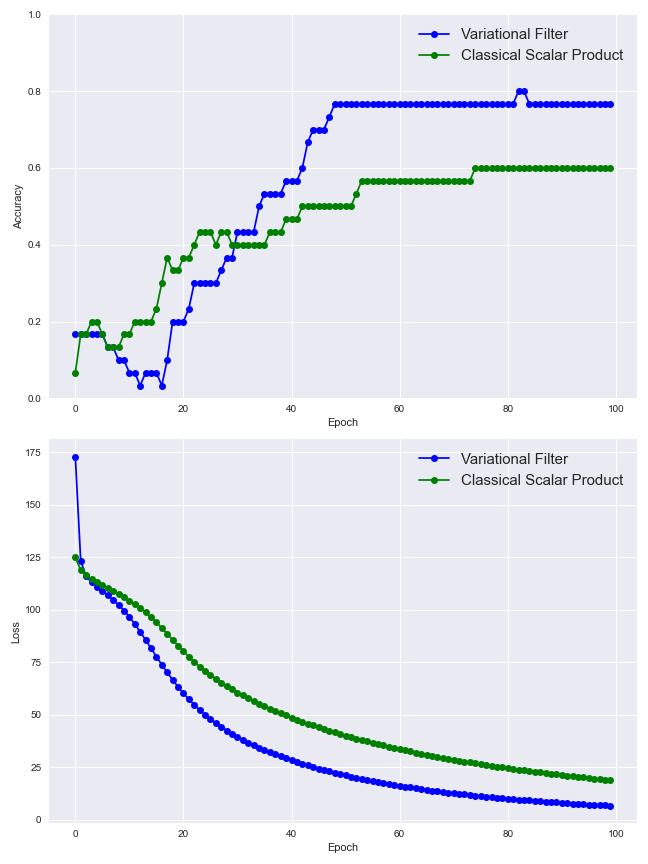}
    
    \caption{The QFilter results as quantum variational with 100 epochs.}
    \label{fig:Result_QFilter_100_Variat}
\end{figure}

Figure\eqref{fig:Result_QFilter_50_Variat} to Fig.\eqref{fig:Result_QFilter_100_Variat} show the QFilter results as quantum variational with 50, 60 and 100 epochs.

\begin{figure}[!ht]
    \centering
    \includegraphics [width=0.4\textwidth]{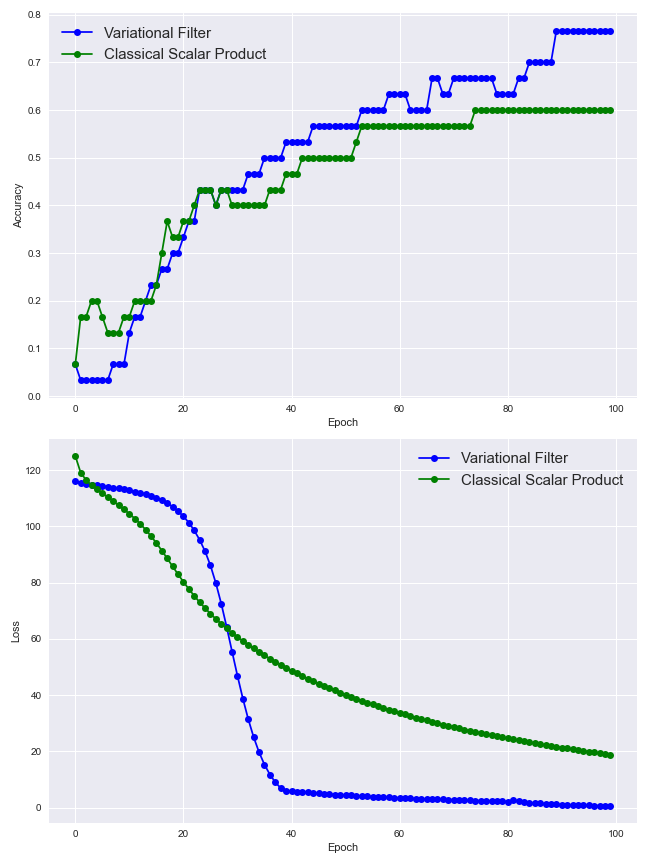}
    \caption{The QFilter results as quantum variational, 2 filters and with 100 epochs.}
    \label{fig:Result_QFilter_Variational_2Filt_100}
\end{figure}

\section{Discussion}
Apart from the results, we want to highlight the points that we consider of the utmost importance in this proof of concept. It is worth remembering that one of the goals of this proof of concept is to offer an approach to using hybrid convolutional filters by harnessing the full potential of current convolutional networks and embedding the quantum part only by changing the scalar product.
Based on the experiments we did, we found it essential to define the following guidelines to achieve the code's adequate scalability.
\begin{enumerate}
    \item Spin up larger jupyter notebook instances, for example the type ml.c5.2xlarge (8 vCPU, 16 GiB Memory) to speed up CPU optimization and quantum simulation time.
    \item Compare the computation time of remote/local simulator.
    \item Increase the number of qubits (filters of size 4x4 and 6x6).
    \item Batch parallelization of the quantum circuits at the gradient and convolution translation operation levels.
\end{enumerate}

By increasing the filter window size, we are simulating more qubits, therefore simulation time also increases exponentially with the number of qubits, at least in full wave-function or state vectors simulators. The gradient computation involves many quantum circuits executions,it scales approximately as $2\times m$ where $m$ is the number of trainable parameters, due to the parameter shift rule used to calculate the trainable quantum filter gradients that propagate through the network, and that really blows up the running time, although we expect high performing remote simulators that are able to batch/parallelize quantum circuits to help. Here is a benchmarking for a fixed set of hyper-parameters of the 16-qubits quantum filter, to get a sense of the running time.
Another aspect to consider is network latency between the CPU and remote QPU (high performance simulators), this can quickly be the major overhead as the feedback loop between classical and quantum computation is iterated many, circuit device executions that are sent to cloud based instances of QPU and simulators must be minimized. For that, the geographical localization of both processors is a must.

The code Ref.\cite{QFilter_code} has been prepared to work for large filter size, so the potential of the API (PennyLane with Amazon Braket) can be exploited in order to parallelize the execution of the Qfilter.

\subsection{Benchmark}
\begin{figure}[!ht]
    \centering
    \includegraphics [width=0.4\textwidth]{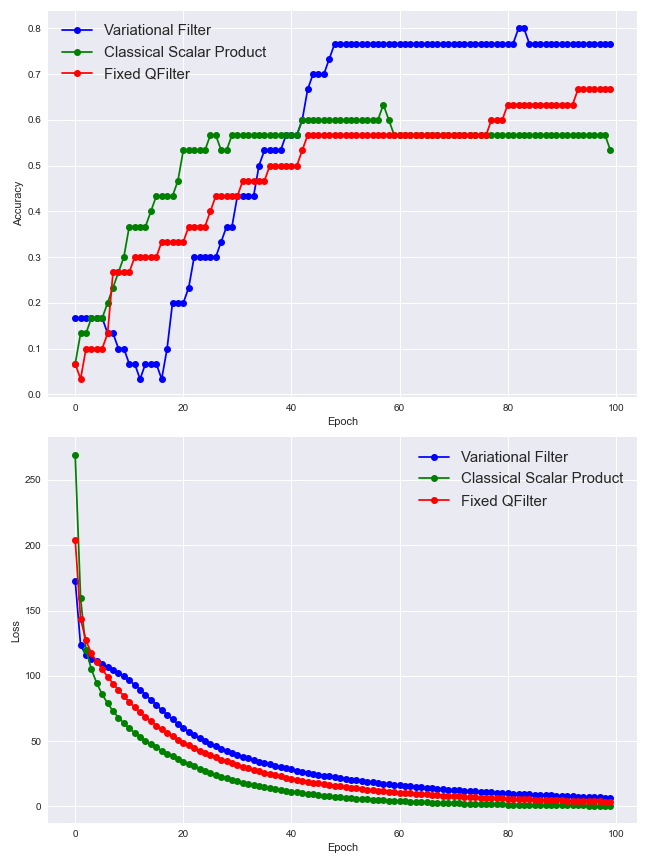}
    \caption{The blue graph marks the precision of our model throughout the training process, the red graph refers to the learning of the quantum fixed model, while the green one is the classical scalar product.}
    \label{fig:Bench_Filters_}
\end{figure}

To test our algorithm, we decided to compare with a similar case already studied Ref.\cite{henderson2019quanvolutional}. In this case, the MNIST is also used, and a fixed quantum filter is applied; that is, it does not train any parameter. The reason for taking random parameters is defined in Ref.\cite{henderson2019quanvolutional}. It is detailed that this type of filters is suitable for detecting vertices, particularly in this dataset; this becomes a relevant skill.
As we can observe Fig.\eqref{fig:Bench_Filters_}, both models end up converging around a $ 40\% $ precision, which, as we have said before, given the small number of images with which we are training, is a successful result. We can also observe in detail the comparison between the variational, fixed and classical filter. It can be seen the best behaviour of our variational filter.

\begin{figure}[htbp]
    \centering
    \includegraphics [width=0.4\textwidth]{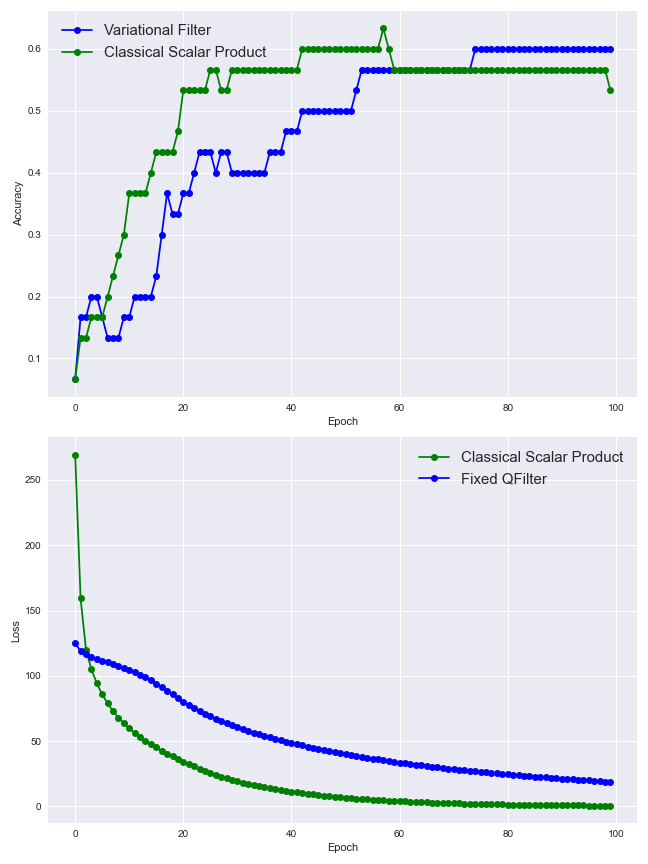}
    \caption{The behaviour’s figure of the QFilter applied to the Fashion MNIST database \cite{xiao2017fashionmnist}. With 50 training images and 30 test images. It is seen that the variational filter continues to behave better. We can observe how for a few epoch, the classical filter behaves better, but for epoch higher than 50, the variational filter exceeds its classical counterpart accuracy.}
    \label{fig:QF_2x2_FMNist_100}
\end{figure}

Also, to attest to the proper functioning of the QFilter, we test it with the MNIST fashion dataset Fig.\eqref{fig:QF_2x2_FMNist_100}. The results have been satisfactory considering the characteristics of said database. We also did the tests with this same dataset, but with the filter fixed to observe its behaviour, and we noticed the variational filter's importance. Since with the fixed filter, it is possible to detect the contours acceptably, in the case of clothing images, as is the point of fashion MNIST, its limitation is seen.

\begin{table}[htbp]
\centering
\begin{tabular}{c|c|cc|c}
 \space &  MacBookPro \cite{MacBook} & \multicolumn{2}{c|}{Amazon Braket\cite{AWS_Braket_instances} } \\
 QFilters & 8-Core Intel & ml.t3.medium  & ml.c5.2xlarge  & \\ \hline 
 2x2 & 3.52 & 19.30 & 4.52  \\
 4x4 & 140.05 & 229.30 & 170.52  \\
 5x6 & -- & -- & --  \\
\end{tabular}
\caption{Table that highlights the importance of latency, the type of instance on which we run our project. This table summarizes in some way the simulation and comparison strategy to take into account for future benchmarks.
In addition to all this, the device to be used must be taken into account. All the values are in second and are the processing image time for the variational filter.}
\label{tab:Benchmark_table_4x4}
\end{table}
One aspect worth finding out is that the accuracy is better for the quantum filter even when the cost function is not. This is an essential advantage because it could mean that hards training are not necessary to achieve better results.

We must emphasize that, due to the era in which we are (NISQ) and due to the novel quantum architecture, the variational circuit's simulation is much more expensive than the classical one. Table \eqref{tab:Benchmark_table_4x4} shows us the processing time in seconds of each image, depending on the filter and the device of the quantum algorithm's execution.

\subsection{Conclusion and future directions}
Throughout this proof of concept, we have been able to test different strategies to tackle somewhat larger than usual hybrid programming problems. Although the number of operations grows, the results obtained have been satisfactory given the experiments carried out, keeping in mind the small number of images used for training. For this reason, we consider that this procedure would be competent in situations of great uncertainty. Another way to continue exploring is the parallelization of the classical part in order to reduce time and study the behaviour in larger images.

\subsection{acknowledgements}
We want to thank Yoshitaka Haribara, Cedric Lin, Pablo Nicolás Nuñez Pölcher and Ricardo García for the support and discussions during this proof-of-concept development. We would also like to thank Xanadu/Pennylane and its team for hosting Qhack, for all the effort put into this super event, and all the sponsors involved. Especially Amazon Web Services, to provide access to its resources (S3, Amazon Braket) to carry out part of this project.

\bibliographystyle{unsrturl}
\bibliography{main}

\end{document}